\documentclass[journal]{IEEEtran}
\usepackage[dvipsnames]{xcolor}
\usepackage{tikz} 
\usepackage{amsmath,amsfonts}
\usepackage{algorithmic}
\usepackage{algorithm}
\usepackage{array}
\usepackage{textcomp}
\usepackage{stfloats}
\usepackage{url}
\usepackage{verbatim}
\usepackage{graphicx}
\usepackage{cite}
\usepackage{pdfpages}
\usepackage{tikz,mathpazo}  
\usetikzlibrary{shadows,arrows,positioning,shapes.geometric,chains} 
\usetikzlibrary{calc} 
\tikzstyle{straightline} = [line width = 1pt,-]
\usepackage{caption}
\usepackage{subcaption}
\usepackage{authblk}
\usepackage{multirow,booktabs,siunitx} 
\ifCLASSINFOpdf
\else
   \usepackage[dvips]{graphicx}
\fi
\usepackage{url}
\usepackage{hyperref}
\usepackage{graphicx}
\usepackage{amssymb}

\begin{document}

\title{Fast Algorithm for Moving Sound Source}
\newcommand*{\email}[1]{%
    \normalsize\href{mailto:#1}{#1}\par
    }
\author{Dong Yang
        % <-this % stops a space
        %\address{GVoice, IEG, Tencent, China}
        }
\affil{Tencent GVoice\\ \email{ydsc531@mail.ustc.edu.cn}}
\markboth{Journal of \LaTeX\ Class Files, Vol. 14, No. 8, August 2015}
{Shell \MakeLowercase{\textit{et al.}}: Bare Demo of IEEEtran.cls for IEEE Journals}
\maketitle
\begin{abstract}
Modern neural network-based speech processing systems usually need to have reverberation resistance, so the training of such systems requires a large amount of reverberation data. In the process of system training, it is now more inclined to use sampling static systems to simulate dynamic systems, or to supplement data through actually recorded data. However, this cannot fundamentally solve the problem of simulating motion data that conforms to physical laws. Aiming at the core issue of insufficient training data for speech enhancement models in moving scenarios, this paper proposes Yang's motion spatio-temporal sampling reconstruction theory to realize efficient simulation of motion continuous time-varying reverberation. This theory breaks through the limitations of the traditional static Image-Source Method (ISM) in time-varying systems. By decomposing the impulse response of the moving image source into two parts: linear time-invariant modulation and discrete time-varying fractional delay, a moving sound field model conforming to physical laws is established. Based on the band-limited characteristics of motion displacement, a hierarchical sampling strategy is proposed: high sampling rate is used for low-order images to retain details, and low sampling rate is used for high-order images to reduce computational complexity. A fast synthesis architecture is designed to realize real-time simulation. Experiments show that compared with the open-source models, the proposed theory can more accurately restore the amplitude and phase changes in moving scenarios, solving the industry problem of motion sound source data simulation, and providing high-quality dynamic training data for speech enhancement models.
\end{abstract}
\newcommand{\cnkeywordsname}{keywords}
\newenvironment{cnkeywords}{\noindent{\bfseries \cnkeywordsname}: }{}

\IEEEpeerreviewmaketitle

\section{Introduction}
In the field of real-time speech enhancement, the performance of data-driven neural network models highly depends on the matching degree between training data and real scenarios \cite{schissler2011gsound}. As a core physical characteristic of the acoustic environment, the simulation quality of reverberation directly affects the robustness of the model. Existing studies mainly focus on static reverberation simulation, approximating the Room Impulse Response (RIR) of fixed scenarios through the Image-Source Method (ISM) \cite{fu2022uformer}. However, in real-time interactive scenarios such as games, dynamic factors such as players' position movement and device attitude changes are common. Static data is difficult to characterize the time-varying sound field characteristics, leading to problems such as speech distortion and tracking failure of the model in moving scenarios \cite{diaz2021gpurir}.
Dynamic reverberation simulation faces dual challenges: first, the motion system is a Linear Time-Varying (LTV) system, which does not satisfy the convolution rules of the traditional Linear Time-Invariant (LTI) system. Direct application of static ISM will lead to distortion; second, the method of fully sampling trajectory points RIR and then splicing signal points, such as the open-source models GSound \cite{schissler2011gsound} and gpuRIR \cite{diaz2021gpurir}, on the one hand, the computational complexity increases with the number of spatial trajectory sampling points, which is difficult to meet the real-time requirements, and on the other hand, there are defects such as phase discontinuity and gain jitter.
To this end, this paper proposes Yang's motion spatio-temporal sampling reconstruction theory. By redefining the image source method for time-varying systems, combined with discrete time-varying fractional delay and hierarchical sampling strategy, the balance between physical authenticity and computational efficiency is achieved. This theory provides a systematic solution for motion sound source data simulation, helping speech enhancement models cope with real-world dynamic scenarios.
\section{Method}
\subsection{Overall Framework}
In static reverberation environments, the Image-Source Method (ISM) is commonly used to approximate the reverberation of time-invariant rooms. The problem we want to solve is a time-varying system, which does not satisfy the operation rules of linear time-invariant systems, so there is no so-called definition of time-varying convolution points. Most engineers and scholars are stuck in the mindset of linear time-invariant systems. In fact, they have been trying to use time-invariant theories to approximate a time-varying system\cite{schissler2011gsound} \cite{diaz2021gpurir}, ignoring the continuous time-varying physical nature of moving objects.
To solve this problem, we must start from the essence of the problem. The essence of ISM is applied in an unbounded space. A unit point source at $\left ( r^\prime,t^\prime \right )$ excites a sound field that propagates as a spherical wave, expressed by the sound field Green's function as:  $g\left(r-r',t-t'\right)=\frac{\delta\left(t-t'-\frac{R}{c}\right)}{4{\pi}}$
, where $t^\prime$ is the sound source excitation time, $r^\prime$ is the sound source position, and $R$ is the distance from the field point to the source point. A complex sound field wave equation is transformed into the superposition of multiple image sources in the free field \cite{allen1979image}. In a static system, this superposition process becomes very simple, that is, the weighted superposition of multiple sound source Dirac delay functions \cite{diaz2021gpurir}.
\begin{equation}
h\left(t\right)=\sum_{i\in\mathcal{N}}{A_i\delta}\left(t-\tau_i\right)
\label{eq:rir}
\end{equation} \par
\begin{figure}[htbp]
            %\centering
            \centerline{\includegraphics[width=0.8\linewidth]{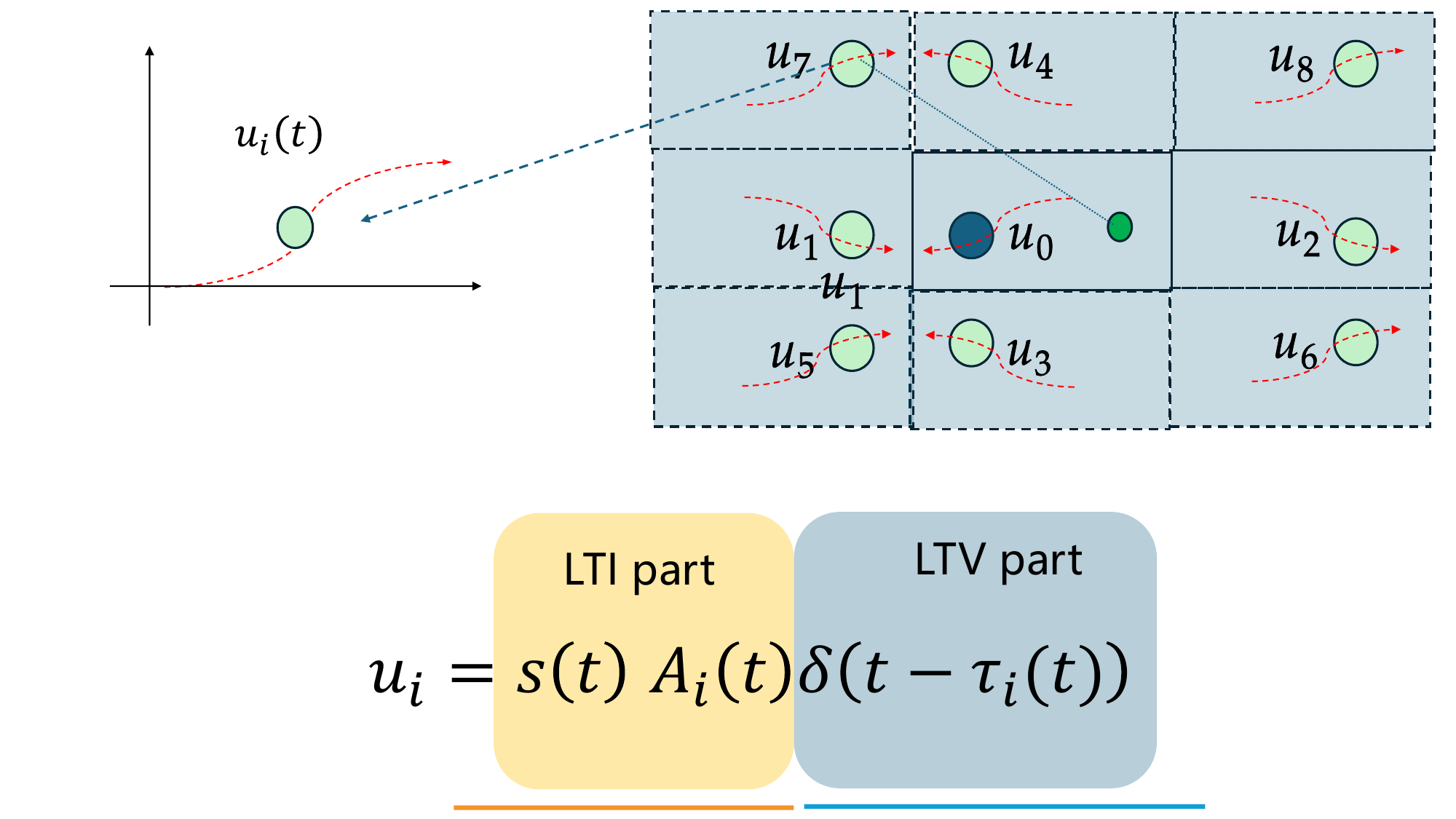}}
            \caption{Schematic diagram of the impulse response process of moving images}
            \label{fig:singleimagemoveshow}
        \end{figure}

As shown in Table. \ref{tab:theory_problem_solution}, in static scenarios, static ISM relies on Linear Time-Invariant (LTI) system theory, and clearly defines the convolution processing rules of signals through static impulse response through $s(t) \circledast h\left(t\right)$. However, in moving scenarios, the impulse response $h(t)$ of motion ISM changes dynamically both with time $t$ and the motion position $p(t)$, making the model seems to be $s(t) \circledast h\left( p(t), t\right)$. However,  the adaptability of the original static formula questionable, because the system changes from LTI to Linear Time-Varying (LTV) system, which does not satisfy the traditional convolution operation rules.
% 定义表格列格式,m{宽度} 用于垂直居中,这里根据内容大致调整宽度
\begin{table}[htbp]
    \centering
    % 使用m列类型实现垂直居中，配合centering实现水平居中
    \begin{tabular}{|>{\centering\arraybackslash}m{0.145\textwidth}|
                     >{\centering\arraybackslash}m{0.145\textwidth}|
                     >{\centering\arraybackslash}m{0.12\textwidth}|} 
        \hline
        \textbf{Existing Theory} & \textbf{Pending Problem} & \textbf{Solution} \\
        \hline
        Static ISM & Motion ISM & Redefined ISM \\ 
        \hline
        $\displaystyle s(t) = s(t) \circledast h$ $\displaystyle h = \sum_{i \in \mathcal{N}} A_i \delta(t - \tau_i)$ & $\displaystyle s(t) = s(t) \circledast h(p(t), t)$? \textcolor{red}{LTV does not satisfy LTI system convolution rules}& Proposed motion spatio-temporal sampling reconstruction theory \\ 
        
        \hline
    \end{tabular}
    \caption{Current Theories, Pending Problems and Solutions}
    \label{tab:theory_problem_solution}
\end{table}%形成核心技术卡点为突破瓶颈,提出重新定义 ISM 方案,以 “杨氏运动时空采样重建理论” 为支撑,重构 ISM 模型适配运动场景,通过创新理论框架,解决动态冲击响应下的计算逻辑与规则兼容问题,为运动场景下的信号处理提供技术突破方向,完整呈现从理论局限到创新解法的科研推进路径 .
Taking this as a starting point, we proceed from the primordial form of the problem, as shown in Fig. \ref{fig:singleimagemoveshow}, we can see that each image system in the motion system process is still independent, and we can still decompose this problem into the motion process of the image source in a single image room. The components of each sound source component at the microphone can be described by an expression. Let $u_i\left(t\right)$ represent the impulse response of a single image source during motion. The process of all image sources propagating to the microphone is independent, so the superposition principle can be used with confidence. Among them, $u_i\left(t\right)$ can be decomposed into two parts: one is the linear time-invariant part $A_i\left(t\right)$, representing the attenuation modulation of the source signal $s\left(t\right)$ during the propagation of energy in space. The other is the time-varying system part $\delta\left(t-\tau_i(t)\right)$ caused by the motion process.

\begin{figure}[htbp]
            \centering
            \centerline{\includegraphics[width=1.0\linewidth]{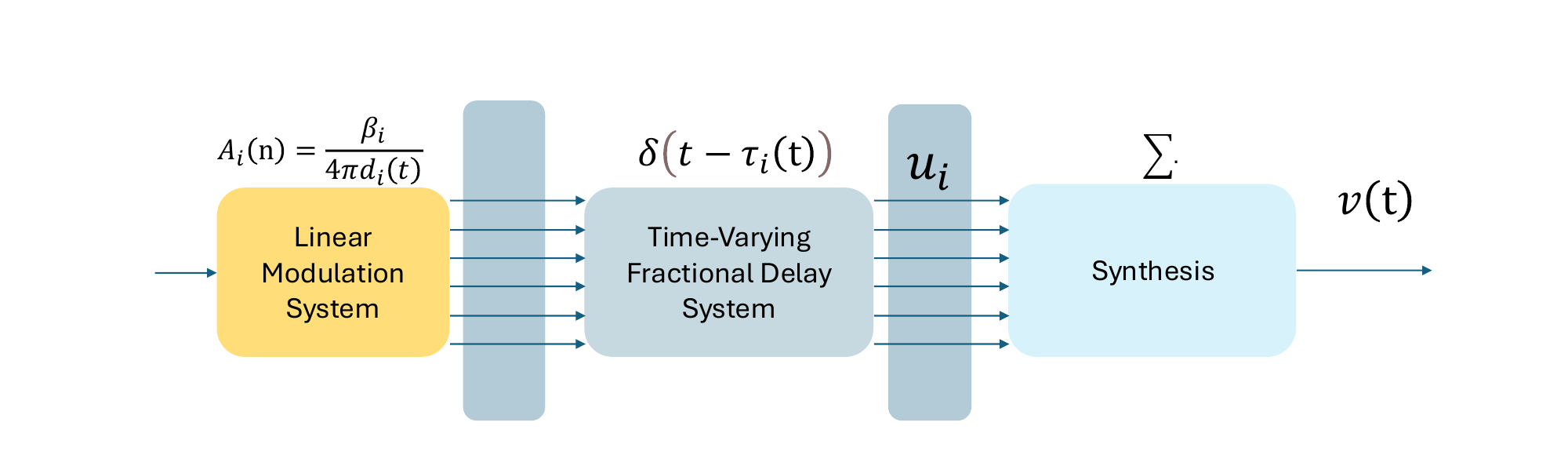}}
            \caption{Computation flow of moving image source synthesis system}
            \label{fig:singleimagemove}
            \end{figure}

Thus, the image source method in the motion process is defined as:
\begin{equation}
v\left(t\right)=\sum_{i\in\mathcal{N}} u_i=\sum_{i\in\mathcal{N}}{s\left(t\right)A_i\left(t\right)}\delta\left(t-\tau_i(t)\right)
\label{eq:mvimg}
\end{equation}
The calculation process is shown in \ref{fig:singleimagemove}.
Where $A_i\left(t\right)=\frac{\beta_i}{4\pi d_i\left(t\right)}$, $d_i\left(t\right)$ is the Euclidean distance from the $i$ image source to the sound pickup, and $\beta_i$ is the reflection attenuation factor \cite{diaz2021gpurir}. Then, the problem is transformed into solving the attenuation modulation and time-varying delay of each sound source. So far, we have analyzed the process of a continuous time-varying system. Next, we will discuss the discretization of the algorithm.
\subsection{Discrete Time-Varying Fractional Delay System}
In digital signal processing, integer delays can be achieved by simple shifting, but fractional delays need to be approximated by filters. The frequency response of an ideal fractional delay filter is $H_d=e^{-j\omega\tau}$ , and the corresponding time-domain impulse response is  $h_d=\frac{sin(\pi\left(n-\tau\right))}{\pi\left(n-\tau\right)}$. This is an infinite-length sequence, which needs to be truncated in practice and an FIR filter is designed for approximation. However, it is impossible to adjust the delay point by point in actual operation. The core idea of the Farrow structure \cite{dooley1999explicit} is to use Horne's rule to express the coefficients of the fractional delay filter as a polynomial function of the delay amount $\tau$. For $N$-th order polynomial approximation, the filter coefficients can be expressed as:
\begin{equation}
h\left(n,\tau\right)=\sum_{k=0}^{M}{c_k\left(n\right)}\tau^k
\label{eq:farrow}
\end{equation}
Among them, $c_k\left(n\right)$ is a fixed coefficient independent of $\tau$, which is only related to the filter order or polynomial order. Generally, first to fourth-order polynomials are used. The higher the order, the higher the accuracy in approximating the ideal delay. The design method of coefficient $c_k\left(n\right)$ is generally optimized based on the frequency domain response, such as complex domain Generalized Least Squares (GLS) approximation of the frequency response. This parameterized representation allows real-time adjustment of the delay by changing $\tau^k$ during operation without recalculating the entire filter coefficients, thus decoupling the time-domain convolution and fractional delay operations, and realizing time-varying fractional delay point by point for each sample with one convolution. Modify the above formula:
\begin{equation}
\label{eq:farrowmv}
h\left(q,\tau\right)=\sum_{k=0}^{M}{c_k\left(q\right)}{\tau_i\left(n\right)}^k
\end{equation}	
Reconstruct the system input and output using Horne's rule:
\begin{equation}
\label{eq:farrowmv2}
 y\left(n\right) = \sum_{k=0}^{M}\left( x\left(n\right)\circledast c_k\right)\tau_i\left(n\right)^k
\end{equation}	
Where, $x\left(n\right)$ is the input signal and, $y\left(n\right)$ is the output signal.

\begin{figure}[htbp]
            \centering
            \centerline{\includegraphics[width=1.0\linewidth]{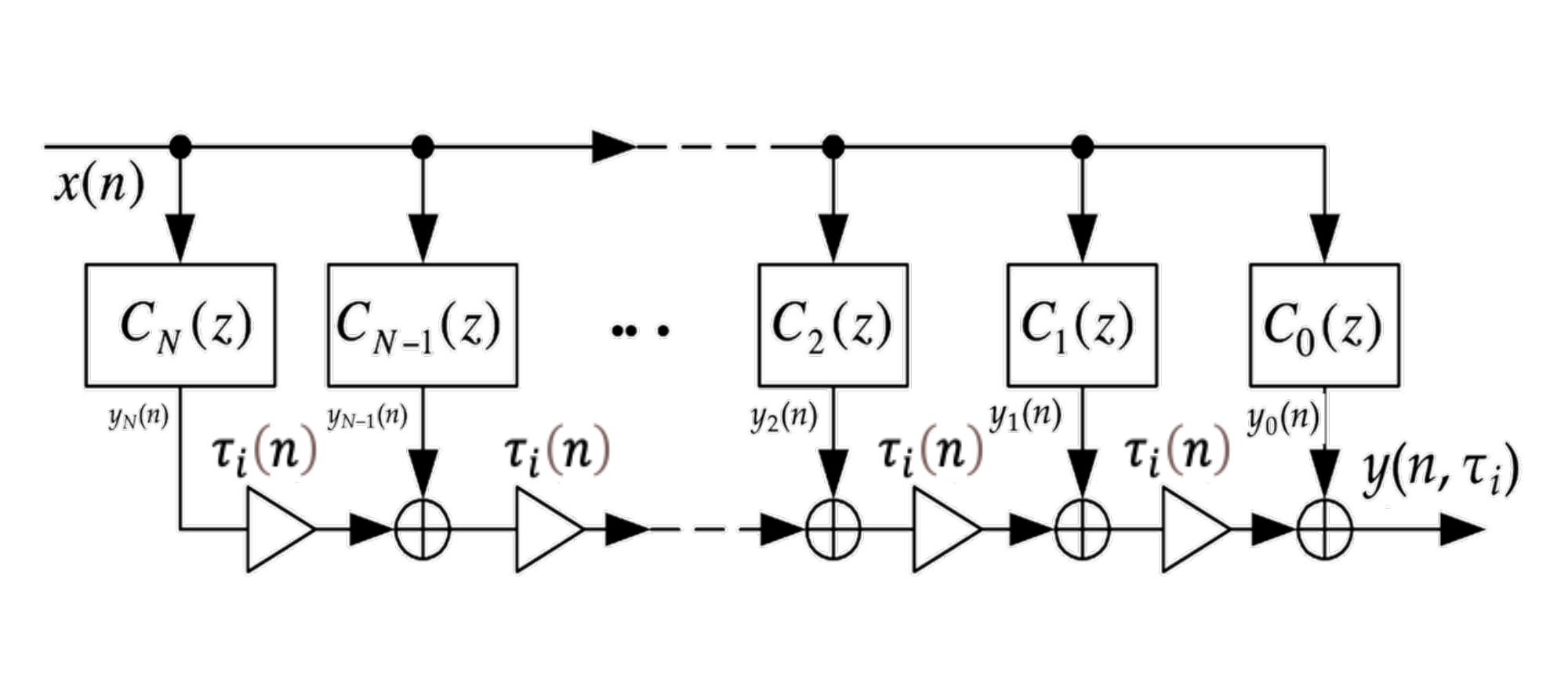}}
            \caption{Discrete time-varying fractional delay system with Farrow architecture}
            \label{fig:farrowmv}
            \end{figure}

Using the Farrow architecture, as shown in Fig.\ref{fig:farrowmv}, the modeling of a time-varying fractional delay system can be realized to approximate $\delta\left(t-\tau_i(n)\right)$.
\subsection{ Simplification of Motion Spatio-Temporal Sampling Reconstruction}
Assuming that the speech system works at 16Khz, and spatial sampling rate is consistent with the speech time sampling rate, this means that each image trajectory needs to generate 16000 fractional delays and attenuations per second. In a medium room with a reverberation time (T60) of 0.6 seconds, each sample needs to generate 45000 images $u_i$, so a total of 720M image samples need to be calculated, which is obviously unacceptable. 
To this end, we first analyze the motion displacement. Assuming we have the  acceleration $a_i\left(t\right)$, instantaneous velocity $v_i\left(t\right)$, displacement $p\left(t\right)$ and initial velocity ${v}_{i,0}$ of a certain image. The instantaneous velocity at time $t$ is given by $v_i\left(t\right)={v}_{i,0}+\int_{0}^{t}{a_i\left(\tau\right)d\tau}$, and the displacement at time $t$ is expressed as $p_i\left(t\right)={v}_{i,0}t+\int_{0}^{t}\int_{0}^{t^\prime}a_i(\tau)d\tau dt^\prime$.
Assume $\mathcal{F}\left(a_i\left(t\right)\right)=\mathbb{A}\left(\omega\right)$, where $\mathcal{F}$ is the Fourier transform. According to the properties of the Fourier transform $\mathcal{F}(\int_{0}^{t}\int_{0}^{\prime}a_i(\tau)d\tau)=-\frac{A(\omega)}{{\omega}^2}$, thus $P_i\left(\omega\right)=\mathcal{F}\left(v_i\left(t\right)\right)-\frac{\mathbb{A}\left(\omega\right)}{\omega^2}{=jv}_{i,0}\frac{\delta\left(\omega\right)}{\omega}-\frac{\mathbb{A}\left(\omega\right)}{\omega^2}$. The motion displacement bandwidth depends on the bandwidth of $\mathbb{A}\left(\omega\right)$ (denoted as $B_a$), and it decays rapidly with the square of the motion frequency $\omega$. Consequently, the motion displacement is inherently a band-limited signal.

\subsubsection{ Bandwidth Analysis of $A_i(t)$ }
For convenience of analysis, the sound source is located at point $o$. It is assumed that the sound source moves only along the $x$-direction and is at position $p\left(t\right)=\left(x_i,{y_i},{z_i}\ \right)$, where the distance from the sound source is $L\left(x_i\right)=\sqrt{{x_i}^2+{y_i}^2+{z_i}^2}$. It is assumed that within a very small time interval $\Delta t$, a displacement of $\epsilon(t)$ occus, such that $x_i^\prime=x_i(t+\Delta t)=x_i(t)+\epsilon(t)$, then $\left(x_i^\prime,{y_i},{z_i}\ \right)$ is the adjacent position of $p\left(t\right)$ in the $i$-th image room. We assume that $x_i$ represents a continuous trajectory. Therefore, as $\Delta t \to 0 $, $\epsilon(t)$ also approaches 0. In this case, $\epsilon(t) \approx	 {\Delta_t}*v_i(t)$, so the bandwidth of $\epsilon(t)$ is equal to that of the displacement $p_i(t)$ , which is also a band-limited. Since $A_i\left(t\right)\propto\frac{1}{L\left(x_i\right)}$, the Taylor expansion of $A_i\left(t\right)$ at $\left(x_i^\prime,{y_i},{z_i}\ \right)$ is:
\begin{equation}
\label{eq:talyor1}
\begin{split}
A_i\left(t\right) &\propto{L\left(x_i^\prime\right)}^{-1}-{x_i^\prime L\left(x_i^\prime\right)}^{-3}\ \epsilon \\
+&{\frac{1}{2}\left({{2x}_i^\prime}^2- 
{y_i}^2-{z_i}^2\right)L\left(x_i^\prime\right)}^{-5}\epsilon^2 \\
&-{\frac{1}{2}x_i^\prime\left({{2x}_i^\prime}^2-{{3y}_i}^2-3{z_i}^2\right)L\left(x_i^\prime\right)}^{-7}\epsilon^3+\cdots
\end{split}
\end{equation}
Since $\epsilon(t)\ll1$, the Taylor series converges as long as $\sqrt{{y_i}^2+{z_i}^2}>0$ . As the image order increases, the distance $x_i^\prime$ increases rapidly, and the high-order terms of the Taylor series decay rapidly, resulting in negligible bandwidth generated by the high-order terms. However, when $L\left(x_i\right)$ is close to the sound source position ($L\left(x_i\right)\rightarrow0$), the nonlinear bandwidth generated by the high-order term $A_i(t)$ cannot be ignored. According to the Nyquist sampling theorem, higher sampling rates need to be used for low-order images and direct sound sources to retain details, while for high-order images in motion, due to their limited bandwidth, lower sampling rates can be used for sampling. Usually, a sampling rate twice the displacement bandwidth $P_i\left(\omega\right)$ can achieve perfect reconstruction.

\subsubsection{ Bandwidth Analysis of $d_i(t)$ }
For simplicity of analysis, it is assumed that at time $t$, the image source is at position $\left(x_i,y_i,z_i\right)$ and moves only along the $x$-axis. We define: 
$d_i\left(t\right)=d\left(x_i\right)=L\left(x_i\right)$,
and the Taylor series expansion of $d\left(x_i\right)$ near $x_i^\prime$ in the $i$-th image room is:
\begin{equation}
    \begin{split}
    d\left(x_i\right) &= L\left(x_i^\prime\right)+\frac{x_i^\prime}{L\left(x_i^\prime\right)}\epsilon+\frac{{y_i}^2+{z_i}^2}{2{L\left(x_i^\prime\right)}^3}\epsilon^2\\
    & +\frac{x_i^\prime\left({y_i}^2+{z_i}^2\right)}{2{L\left(x_i^\prime\right)}^5}\epsilon^3+\cdots
    \end{split}
\end{equation}
Similarly, as the image order increases, the distance $x_i^\prime$ increases rapidly, and the high-order terms of the Taylor series decay rapidly, resulting in negligible bandwidth generated by the high-order terms. In this way, we also arrive at a similar conclusion:  since for high-order image sources the high-order terms in Taylor series decay too quickly, in practice, $d\left(x_i\right)$ can be approximated directly by the first-order term as
$d\left(x_i\right)= L\left(x_i^\prime\right)+\frac{x_i^\prime}{L\left(x_i^\prime\right)}\epsilon$
, so the bandwidth of $d\left(x_i\right)$ is basically consistent with the displacement bandwidth, thus a lower sampling rate can be used for sampling. Meanwhile, higher sampling rates need to be used for low-order images and direct sound sources.
\subsubsection{ Fast Architecture for Moving Image Source Synthesis}
\begin{figure*}[htbp]
            \centering
            \centerline{\includegraphics[width=1.0\linewidth]{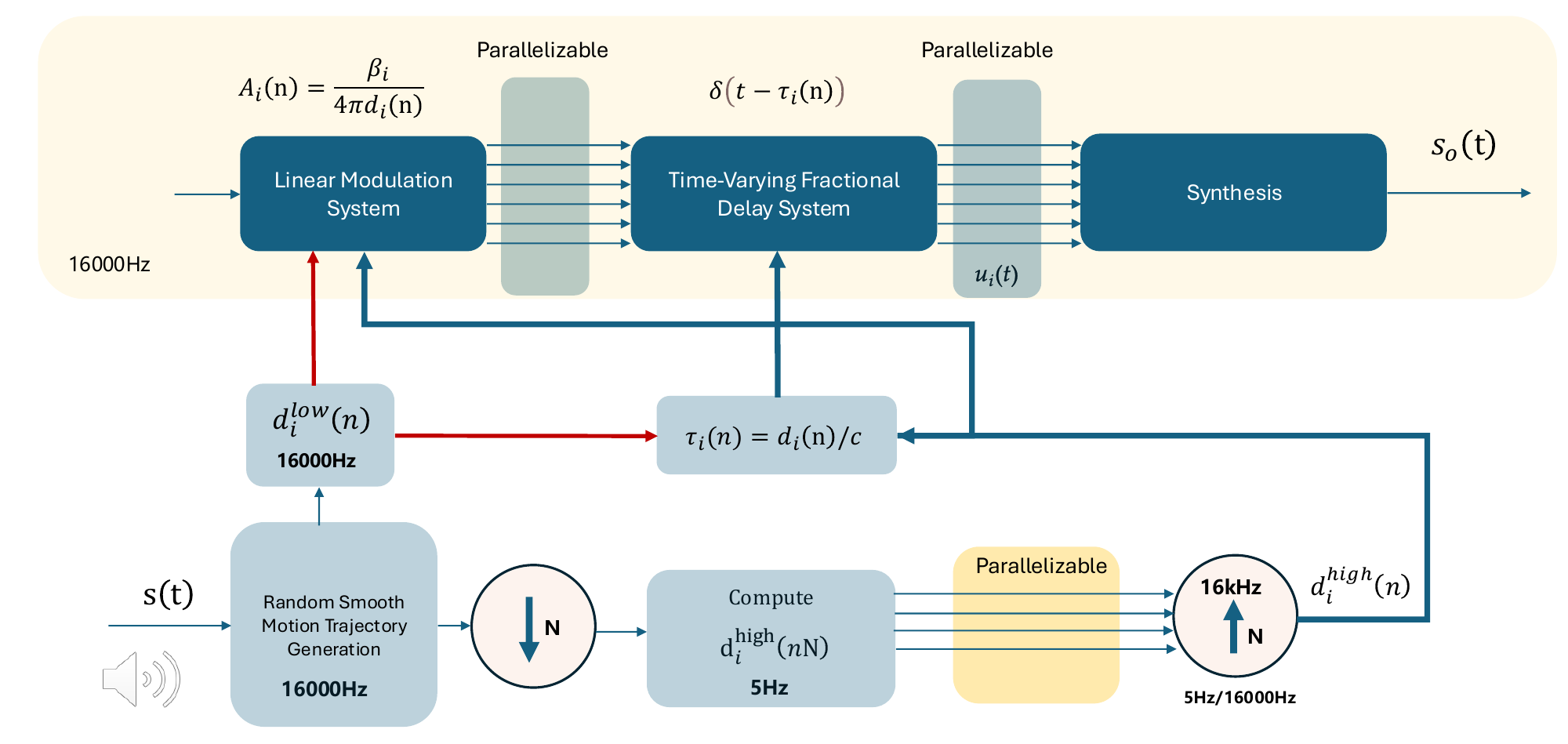}}
            \caption{Computation flow of fast moving image source synthesis system}
            \label{fig:fastalg}
            \end{figure*}

Building on the above theoretical framework, we design the architecture illustrated in \ref{fig:fastalg} with the implementation steps detailed as follows: First, we randomly sample a segment of the motion bandlimited spatiotemporal trajectory $p(n)$. Based on this trajectory, we first compute the lower-order image sources to obtain $d_i^{low}(n)$, which are calculated at a standard spatiotemporal sampling rate (e.g., 16 kHz in this work, consistent with the temporal sampling rate of audio signals). It should be noted that the spatiotemporal sampling rate mentioned herein specifically refers to the sampling frequency on the spatial motion trajectory corresponding to equal-interval time, and is distinct from the time-scale sampling rate used in audio sampling.
Next, we downsample the trajectory by a factor of $N$ and compute the higher-order image sources $d_i^{high}(nN)$ at this reduced spatiotemporal sampling rate. Subsequently, $d_i^{high}(n)$ is recovered using an upsampling algorithm. This approach avoids generating image sources at every individual time step—notably, generating sources at all time steps would lead to prohibitive computational complexity for large reverberation times $T_{60}$, as the complexity of ISM image sources grows exponentially with $T_{60}$. Once $d_i^{low}(n)$ and $d_i^{high}\left(n\right)$, are computed, we merge them to obtain the full set of image sources $d_i(n)$. Using $d_i(n)$, we further calculate the time delays $\tau_i\left(n\right)$, where $\tau_i\left(n\right)=d_i\left(n\right)/c$ and $c$ denotes the speed of sound. Finally, the output signal of the motion time-varying system is obtained via the linear modulation system and the discrete-time time-varying fractional delay system as Eq.\ref{eq:fastalg}.
\begin{equation}
\label{eq:fastalg}
s_o\left(t\right)=\sum_{i\in\mathcal{N}} u_{i}\left(n\right)=\sum_{i\in\mathcal{N}}{s\left(n\right)A_i\left(n\right)}\delta\left(t-\tau_i(n)\right)
\end{equation}

From Fig. \ref{fig:fastalg} and Eq. \ref{eq:fastalg}, we can observe that each image point of the algorithm is independent. This fundamental property renders parallel processing feasible for nearly all stages of the algorithm.

\section{Performance Evaluation}
\begin {figure}[htbp]
\centering
\centerline {\includegraphics [width=1.0\linewidth]{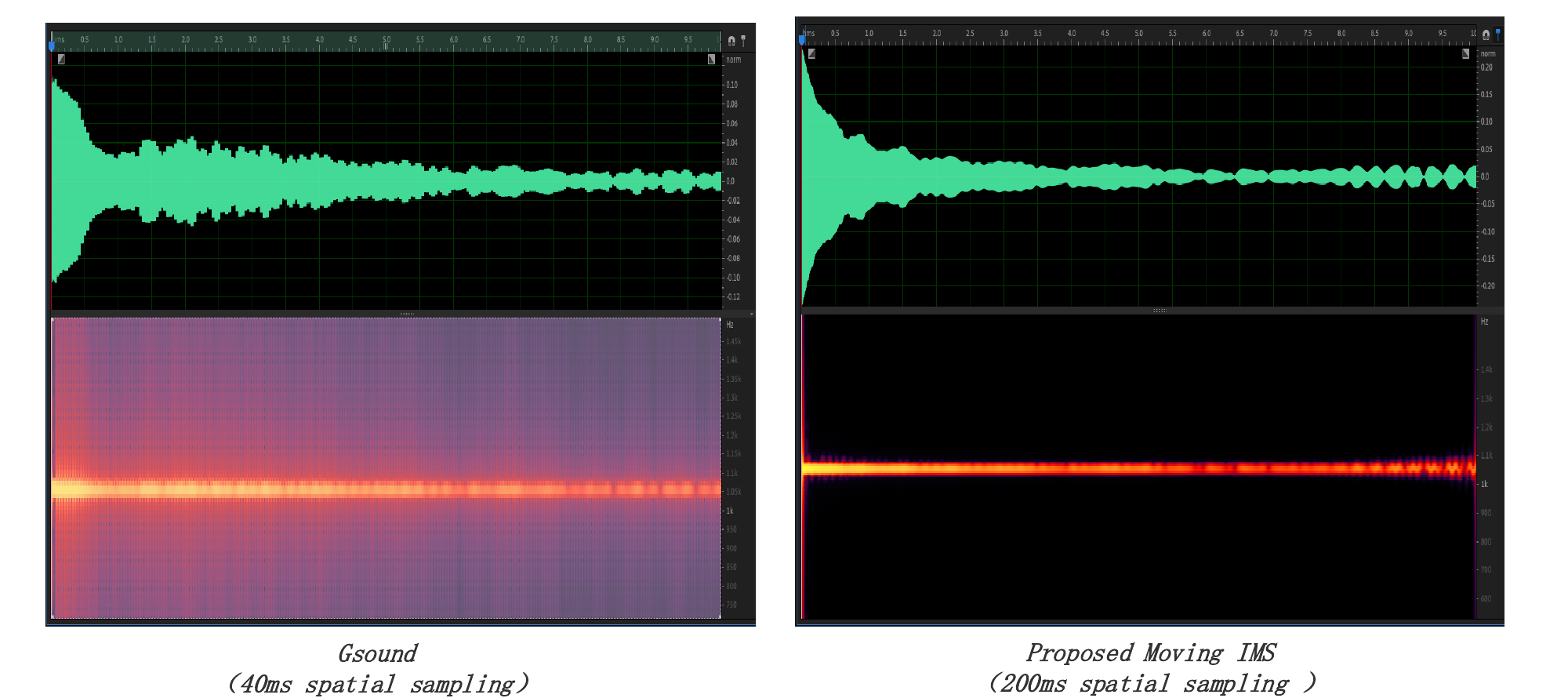}}
\caption {Comparison of synthesis effects of moving sound sources}
\label {fig:results}
\end {figure}

We evaluate the advantages of our algorithm from two dimensions: first, the quality of the generated data; second, the tracking performance of the model on moving targets after incorporating the moving sound source data generated by the algorithm into the training process. Specifically, we analyze the difference in tracking performance of models trained with moving data generated in this paper and those not trained with such data for moving targets in microphone array enhancement scenarios. Taking the well-known open-source baseline model  \cite {schissler2011gsound} as the comparison object, the experiment uses a 1kHz dry sine signal (duration 2 seconds) as the excitation signal, and the sound source moves along a slow uniform curve away from the sound pickup device (experimental results are shown in Fig. \ref {fig:results}). The results show that  has a spatiotemporal  sampling rate of 25Hz (5 times that of our scheme), but its synthesis effect has significant phase discontinuity and gain sawtooth phenomenon;while the proposed algorithm can better restore the amplitude and phase characteristics during sound field changes.
Although open-source models such as GSound try to improve dynamic effects by increasing the spatiotemporal sampling rate, they still cannot overcome defects such as phase discontinuity and gain jitter, making it difficult to accurately restore sound field changes in moving scenarios. Therefore, building a dynamic reverberation data generation framework that balances physical authenticity and computational efficiency has become a core path to break through the robustness bottleneck of neural network models in real dynamic scenarios, which is also the core research value of the motion spatio-temporal sampling reconstruction theory. To demonstrate the effectiveness of this method, we construct a model $\mathcal{F}_{\Theta}$
capable of processing dual-channel microphone signals $x_1$ and $x_2$, that can enhance the speech signal in specific regions through the spatial information of $x=\{x_1,x_2\}$. This model can output the speech estimation signal $\hat{y}$ (the corresponding ground truth is $y$). Specifically,~$\hat{y}=\mathcal{F}_{\Theta}(\mathbf{X_1},\mathbf{X_2})$, where $\mathbf{X_i}=STFT(x_i(t))$ and $\mathbf{X_i}\in \mathbb{C}$ ($STFT$ is the Short-Time Fourier Transform) with $i=\{0,1\}$. ${\mathcal{F}_{\Theta}}$ is a complex model in the time-frequency domain based on UNet \cite{fu2022uformer} and Transformer architectures. In the experiment, the microphone spacing is set to 15 cm, and a dual-channel speech dataset ${\mathcal{D}{({\theta})}}={(x, y)}$ is constructed based on this geometric structure, where $\theta$ denotes the angle between the normal direction of the microphone array's connecting line and the sound source: when $\left\vert\theta\right\vert<g$, the sound source is within the $g$-angle range, corresponding to the sub-dataset ${\mathcal{D}_{in}}$ that we aim to extract this target sound source; when $\left\vert\theta\right\vert>=g$, the sound source is outside the $g$ angle range, corresponding to the sub-dataset ${\mathcal{D}_{out}}$ that we consider it an interfering sound source and aim to suppress it. The dataset ${\mathcal{D}{({\theta})}}$ includes two parts of data: one is dual-channel static reverberation speech data generated by gpuRIR; the other is dynamic motion simulation speech data generated by the method in this paper. In the process of generating dynamic data, motion trajectories are randomly generated, the spatiotemporal sampling rate is consistent with the speech sampling rate (16000Hz), and the spatial downsampling ratio is 3200. The ratio of dynamic data to static data is 1:10. All speech data are randomly mixed with static/dynamic dual-channel noise data (the construction method of noise data is similar to that of speech data). The Loss function is designed as:
\begin{equation}
    \mathcal{L}=dist(\mathcal{F}_{\Theta}({\mathcal{D}_{in}}),\mathbf{y}) +   dist(\mathcal{F}_{\Theta}({\mathcal{D}_{out}}),\mathbf{0})
\end{equation}
Table \ref {tab:algorithm_comparison} compares the performance of models trained with pure static data and models trained with mixed data (static + dynamic) in moving scenarios (the ratio of moving to fixed data in the test data is 1:1). The results show that the model trained with mixed data has significant advantages in three key speech quality indicators: SDR, PESQ-WB, and STOI.
\begin{table}[h!]
\centering
\begin{tabular}{|l|l|l|l|l|}
\hline
\multicolumn{1}{|c|}{\textbf{Test Data}} & \multicolumn{3}{c|}{\textbf{Moving + Fixed Data}} \\ \hline
\textbf{Model} & \textbf{SDR (dB)} & \textbf{PESQ-WB} & \textbf{STOI}  \\ \hline
Before Processing & 2.37 & 1.95 & 0.8504  \\ \hline
Static Data Model & 16.34 & 3.24 & 0.9435  \\ \hline
Mixed Data Model & 18.65 & 3.35 & 0.9738 \\ \hline
\end{tabular}
\caption{Performance comparison between models trained with static data and mixed data in moving scenarios}
\label{tab:algorithm_comparison}
\end{table}
\section{Conclusion}
Aiming at the problem of insufficient training data for speech enhancement models in moving scenarios, this paper proposes a motion spatio-temporal sampling reconstruction theory to realize efficient simulation of motion continuous time-varying reverberation. This theory breaks through the limitations of the traditional static Image-Source Method (IMS) in time-varying systems. By decomposing the impulse response of the moving image source into two parts: linear time-invariant modulation and discrete time-varying fractional delay, a moving sound field model conforming to physical laws is established.
Based on the band-limited characteristics of motion displacement, the proposed hierarchical sampling strategy uses high sampling rates for low-order images to retain details and low sampling rates for high-order images to reduce computational complexity. A fast synthesis architecture combined with the Farrow structure is designed to realize real-time simulation.
Experimental results show that compared with the open-source model, the proposed theory can more accurately restore the amplitude and phase changes in moving scenarios, effectively solving the problem of motion sound source data simulation in the industry. At the same time, the model trained with dynamic data generated by this theory outperforms the model trained only with static data in speech quality indicators such as SDR, PESQ-WB, and STOI, significantly improving the tracking performance and robustness of the multi-channel end-to-end speech enhancement algorithm.

\vfill\pagebreak
%\subsection{References}

%\begin{thebibliography}{1}
\nocite{*}
\bibliographystyle{IEEEtran}
\small\bibliography{IEEEabrv, ref} 

\end{document}